\documentclass[preprintnumbers,amsmath,amssymbm,prd]{revtex4}
\usepackage{epsfig}
\usepackage{graphicx}

\begin{document}
\title{Gauss-Bonnet black holes supporting massive scalar field configurations: The large-mass regime}
\author{Shahar Hod}
\affiliation{The Ruppin Academic Center, Emeq Hefer 40250, Israel}
\affiliation{ }
\affiliation{The Hadassah Institute, Jerusalem 91010, Israel}
\date{\today}

\begin{abstract}
\ \ \ It has recently been demonstrated that black holes with
spatially regular horizons can support external scalar fields
(scalar hairy configurations) which are non-minimally coupled to the
Gauss-Bonnet invariant of the curved spacetime. The composed
black-hole-scalar-field system is characterized by a critical
existence line $\alpha=\alpha(\mu r_{\text{H}})$ which, for a
given mass of the supported scalar field, marks the threshold for
the onset of the spontaneous scalarization phenomenon [here
$\{\alpha,\mu,r_{\text{H}}\}$ are respectively the dimensionless non-minimal 
coupling parameter of the field theory, the proper mass of the
scalar field, and the horizon radius of the central supporting black
hole]. In the present paper we use analytical techniques in order to
explore the physical and mathematical properties of the
marginally-stable composed black-hole-linearized-scalar-field
configurations in the eikonal regime $\mu r_{\text{H}}\gg1$ of large
field masses. In particular, we derive a remarkably compact
analytical formula for the critical existence-line
$\alpha=\alpha(\mu r_{\text{H}})$ of the system which separates bare
Schwarzschild black-hole spacetimes from composed hairy (scalarized)
black-hole-field configurations.
\end{abstract}
\bigskip
\maketitle

\section{Introduction}

The mathematically elegant no-hair theorems presented in
\cite{Bek1,Her1,Sotc} have revealed the physically important fact
that, within the framework of classical general relativity,
spherically symmetric black holes with regular horizons cannot
support external static matter configurations which are made of
scalar fields with minimal coupling to gravity. As explicitly proved
in \cite{BekMay,Hod1}, the intriguing no-hair property of static
black holes can also be extended to the physical regime of scalar
matter fields which are characterized by a non-trivial (non-minimal)
coupling to the Ricci curvature scalar of the corresponding
spherically symmetric spacetimes.

Interestingly, later developments
\cite{Sot5,GB1,GB2,Bab,Her2,ChunHer} have revealed the intriguing
fact that spatially regular hairy matter configurations which are
made of scalar fields with non-minimal couplings to the Gauss-Bonnet
curvature invariant ${\cal G}$ may be supported in curved black-hole
spacetimes. In particular, it has been proved \cite{GB1,GB2,ChunHer}
that, in extended Scalar-Tensor-Gauss-Bonnet theories whose actions
contain a non-trivial field-curvature coupling term of the form
$f(\phi){\cal G}$ \cite{Notephi}, black holes with regular horizons
may support scalar fields with non-trivial spatial profiles (see
\cite{Herr1,Herr2,Hodchar} for the physically related model of
spontaneously scalarized charged black-hole spacetimes which owe their
existence to a non-trivial coupling between the external scalar
field and the electromagnetic field tensor of the central supporting charged black
hole).

In a physically realistic field theory, the {\it spontaneous
scalarization} phenomenon should be characterized by a non-trivial
coupling function $f(\phi)$ whose mathematical form allows the
existence of bare (non-scalarized) black-hole solutions in the
weak-coupling regime \cite{GB1,GB2,ChunHer}. Specifically, the
physically important studies presented in \cite{GB1,GB2,ChunHer}
have considered Scalar-Tensor-Gauss-Bonnet theories whose coupling
functions are characterized by the limiting behavior
$f(\phi\to0)\propto\alpha\phi^2$ in the weak-field regime. Here the
physical parameter $\alpha$ is the dimensionless coupling constant
of the non-trivial field theory [see Eq. (\ref{Eq10}) below].

Intriguingly, it has recently been proved \cite{Macn} that, for
non-minimally coupled {\it massive} scalar fields, the composed
black-hole-field system is characterized by a critical {\it
existence-line} $\alpha=\alpha(\mu r_{\text{H}})$ which separates
bare Schwarzschild black holes from hairy (scalarized)
black-hole-field solutions of the field equations (here $\mu$ is the
proper mass of the supported scalar field and $r_{\text{H}}$ is the
horizon radius of the central black hole). In particular, the
existence-line of the system corresponds to linearized
marginally-stable scalar field configurations which are supported by
central Schwarzschild black holes. [In the physics literature
\cite{Hodlit,Herlit}, the supported linearized scalar field
configurations are usually called scalar `clouds' in order to
distinguish them from self-gravitating (non-linear) hairy matter
configurations]. Interestingly, the numerical results presented in
\cite{Macn} have revealed the fact that, for a given value of the
dimensionless coupling parameter $\alpha$, the horizon radius (mass)
of the central supporting black hole is a monotonically decreasing
function of the mass of the supported scalar field.

The main goal of the present paper is to explore, using 
analytical techniques, the physical and mathematical properties of
the composed
Schwarzschild-black-hole-nonminimally-coupled-linearized-massive-scalar-field
cloudy configurations. In particular, using a WKB analysis in the
dimensionless large-mass $\mu r_{\text{H}}\gg1$ regime, we shall
derive a resonance formula that 
provides a remarkably compact analytical description of the critical 
existence-line $\alpha=\alpha(\mu r_{\text{H}})$ of the composed
Schwarzschild-black-hole-massive-scalar-field system. Interestingly,
the derived resonance formula [see Eq. (\ref{Eq24}) below] would provide a simple {\it analytical} 
explanation for the {\it numerically} observed \cite{Macn} monotonic
behavior of the function $r_{\text{H}}=r_{\text{H}}(\mu;\alpha)$
along the critical existence-line of the system.

\section{Description of the system}

We shall study analytically the discrete resonant spectrum which
characterizes the composed
Schwarzschild-black-hole-linearized-massive-scalar-field
configurations in the physical regime of large field masses. As
shown numerically in \cite{GB1,GB2,ChunHer,Macn}, the spatially regular
cloudy field configurations owe their existence to their non-trivial
coupling to the Gauss-Bonnet invariant ${\cal G}\equiv
R_{\mu\nu\rho\sigma}R^{\mu\nu\rho\sigma}-4R_{\mu\nu}R^{\mu\nu}+R^2$
of the curved spacetime. The black-hole spacetime is characterized
by the spherically-symmetric curved line element \cite{Noteunits}
\begin{equation}\label{Eq1}
ds^2=-h(r)dt^2+{1\over{h(r)}}dr^2+r^2(d\theta^2+\sin^2\theta
d\phi^2)\  ,
\end{equation}
where
\begin{equation}\label{Eq2}
h(r)=1-{{r_{\text{H}}}\over{r}}\  .
\end{equation}
Here $r_{\text{H}}=2M$ is the horizon radius of the central
supporting Schwarzschild black hole of mass $M$.

The composed black-hole-field system is characterized by the action
\cite{GB1,GB2,Macn,Noteho}
\begin{equation}\label{Eq3}
S={1\over2}\int
d^4x\sqrt{-g}\Big[R-{1\over2}\nabla_{\alpha}\phi\nabla^{\alpha}\phi-{1\over2}\mu^2\phi^2+f(\phi){\cal
G}\Big]\  ,
\end{equation}
where the radius-dependent Gauss-Bonnet curvature invariant of the
Schwarzschild black-hole spacetime is given by
\begin{equation}\label{Eq4}
{\cal G}={{12r^2_{\text{H}}}\over{r^6}}\  .
\end{equation}
The scalar function $f(\phi)$ in (\ref{Eq3}) controls the
non-minimal coupling between the Gauss-Bonnet invariant of the
curved spacetime and the massive scalar field. As shown in
\cite{GB1,GB2,Macn}, in order to guarantee the existence of bald
(non-scalarized) black-hole solutions in the field theory, this coupling
function should have the universal leading-order quadratic behavior
\begin{equation}\label{Eq5}
f(\phi)={1\over8}\eta\phi^2\
\end{equation}
in the linearized regime. The physical parameter $\eta$, which
controls the strength of the non-trivial quadratic coupling between
the massive scalar field and the Gauss-Bonnet curvature invariant,
has the dimensions of length$^2$.

Using the functional expression \cite{Notelm}
\begin{equation}\label{Eq6}
\phi(r,\theta,\phi)=\sum_{lm}{{\psi_{lm}(r)}\over{r}}Y_{lm}(\theta)e^{im\phi}\
\end{equation}
for the non-minimally coupled static scalar field and defining the
tortoise radial coordinate $y$ by the relation \cite{Notemap}
\begin{equation}\label{Eq7}
{{dr}\over{dy}}=h(r)\  ,
\end{equation}
one finds that the spatial behavior of the supported massive scalar
field configurations in the Schwarzschild black-hole spacetime
(\ref{Eq1}) is determined by the Schr\"odinger-like ordinary
differential equation \cite{GB1,GB2,Macn}
\begin{equation}\label{Eq8}
{{d^2\psi}\over{dy^2}}-V\psi=0\  ,
\end{equation}
where \cite{GB1,GB2,Macn}
\begin{equation}\label{Eq9}
V(r)=\Big(1-{{r_{\text{H}}}\over{r}}\Big)\Big[{{l(l+1)}\over{r^2}}+{{r_{\text{H}}}\over{r^3}}+\mu^2-{{\alpha
r^4_{\text{H}}}\over{r^6}}\Big]\  .
\end{equation}
Here
\begin{equation}\label{Eq10}
\alpha\equiv 3\eta r^2_{\text{H}}\
\end{equation}
is the dimensionless non-trivial coupling parameter of the composed
black-hole-massive-scalar-field system.

The Schr\"odinger-like equation (\ref{Eq8}) with its effective
radial potential (\ref{Eq9}), supplemented by the physically
motivated boundary conditions of exponentially decaying scalar
eigenfunctions at spatial infinity and a spatially regular
functional behavior at the black-hole horizon \cite{GB1,GB2,Macn},
\begin{equation}\label{Eq11}
\psi(r\to\infty)\sim r^{-1}e^{-\mu r}\to 0\ \ \ \ ; \ \ \ \
\psi(r=r_{\text{H}})<\infty\ ,
\end{equation}
determine the discrete resonant spectrum
$\{\alpha_n(\mu,r_{\text{H}})\}_{n=0}^{n=\infty}$ which
characterizes the composed cloudy
black-hole-nonminimally-coupled-linearized-massive-scalar-field
configurations. In particular, the fundamental resonant mode, 
$\alpha_0=\alpha_0(\mu,r_{\text{H}})$, determines the critical
existence-line of the field theory in the curved black-hole
spacetime.

\section{The discrete resonant spectrum of the composed
black-hole-linearized-massive-scalar-field system: A WKB analysis}

In the present section we shall use analytical techniques in order
to study the discrete resonant spectrum
$\{\alpha_n(\mu,r_{\text{H}})\}_{n=0}^{n=\infty}$ of the
dimensionless scalar-Gauss-Bonnet coupling parameter which
characterizes the composed
black-hole-linearized-massive-scalar-field configurations in the
large-mass regime
\begin{equation}\label{Eq12}
\mu r_{\text{H}}\gg \text{max}\{1,l\}\  .
\end{equation}

We first point out that, in terms of the tortoise coordinate $y$
[see Eq. (\ref{Eq7})], the Schr\"odinger-like radial differential
equation (\ref{Eq8}) has a mathematical form which is amenable to a
standard WKB analysis. In particular, a standard second-order WKB
analysis for the spatially regular bound-state resonances of the
Schr\"odinger-like ordinary differential equation (\ref{Eq7}) yields
the well-known quantization condition \cite{WKB1,WKB2,WKB3,Notephas}
\begin{equation}\label{Eq13}
\int_{y_-}^{y_+}dy\sqrt{-V(y;\alpha)}=\big(n-{1\over4}\big)\cdot\pi\
\ \ \ ; \ \ \ \ n=1,2,3,...\  .
\end{equation}
The two boundaries $\{y_-,y_+\}$ of the WKB integral relation
(\ref{Eq13}) are determined by the classical turning points of the
radial binding potential (\ref{Eq9}) [that is, $V(y_-)=V(y_+)=0$].
The integer $n$ is the resonance parameter which characterizes the
discrete bound-state resonant modes of the composed
black-hole-nonminimally-coupled-massive-scalar-field system.

Taking cognizance of the differential relation (\ref{Eq7}), one can
express the WKB integral relation (\ref{Eq13}), which characterizes
the composed black-hole-linearized-massive-scalar-field
configurations, in the form
\begin{equation}\label{Eq14}
\int_{r_-}^{r_+}dr{{\sqrt{-V(r;\alpha)}}\over{h(r)}}=\big(n-{1\over4}\big)\cdot\pi\
\ \ \ ; \ \ \ \ n=1,2,3,...\  .
\end{equation}
The radial turning points $\{r_-,r_+\}$ of the binding potential
(\ref{Eq9}) are determined by the two polynomial relations
\begin{equation}\label{Eq15}
1-{{r_{\text{H}}}\over{r_-}}=0\
\end{equation}
and
\begin{equation}\label{Eq16}
{{l(l+1)}\over{r^2_+}}+{{r_{\text{H}}}\over{r^3_+}}+\mu^2-{{\alpha
r^4_{\text{H}}}\over{r^6_+}}=0\  .
\end{equation}

As we shall now show explicitly, the WKB integral relation
(\ref{Eq14}) can be studied {\it analytically} in the large-mass
regime (\ref{Eq12}). In particular, defining the dimensionless
radial coordinate
\begin{equation}\label{Eq17}
x\equiv {{r-r_{\text{H}}}\over{r_{\text{H}}}}\  ,
\end{equation}
one can expand the effective binding potential of the composed
black-hole-massive-field system in the form
\begin{equation}\label{Eq18}
V[x(r)]=-\Big({{\alpha}\over{r^2_{\text{H}}}}-\mu^2\Big)\cdot x +
\Big({{7\alpha}\over{r^2_{\text{H}}}}-\mu^2\Big)\cdot x^2+O(x^3)\  .
\end{equation}

The near-horizon radial potential (\ref{Eq18}) has the form of an
effective binding potential. In particular, from (\ref{Eq18}) one
finds that the two turning points $\{x_-,x_+\}$ of the WKB integral
relation (\ref{Eq14}) are given by the simple dimensionless
functional expressions
\begin{equation}\label{Eq19}
x_-=0\
\end{equation}
and
\begin{equation}\label{Eq20}
x_+={{{{\alpha}\over{r^2_{\text{H}}}}-\mu^2}\over{{{7\alpha}\over{r^2_{\text{H}}}}-\mu^2}}\
.
\end{equation}

Taking cognizance of Eqs. (\ref{Eq17}), (\ref{Eq18}), (\ref{Eq19}),
and (\ref{Eq20}), one finds that, in the large-mass regime
(\ref{Eq12}), the WKB integral equation (\ref{Eq14}) can be
approximated by \cite{Notexp}
\begin{equation}\label{Eq21}
\sqrt{\alpha-\mu^2r^2_{\text{H}}}\int_{0}^{x_+}dx
\sqrt{{{1}\over{x}}-{{1}\over{x_+}}}=\big(n-{1\over4})\cdot\pi\ \ \
\ ; \ \ \ \ n=1,2,3,...\ .
\end{equation}
Interestingly, and most importantly for our analysis, the integral
on the l.h.s of Eq. (\ref{Eq21}) can be evaluated analytically to
yield the WKB resonance relation
\begin{equation}\label{Eq22}
{{\alpha-\mu^2r^2_{\text{H}}}\over{2\sqrt{7\alpha-\mu^2r^2_{\text{H}}}}}=n-{1\over4}\
\ \ \ ; \ \ \ \ n=1,2,3,...\ .
\end{equation}
The solution of the polynomial equation (\ref{Eq22}) for the
dimensionless coupling parameter $\alpha$ of the composed
black-hole-massive-field theory is given by the rather cumbersome
expression
\begin{equation}\label{Eq23}
\alpha_n=\mu^2 r^2_{\text{H}}+14(n+{3\over4})^2+2\sqrt{6\mu^2 r^2_{\text{H}}(n+{3\over4})^2+49(n+{3\over4})^4}\ \ \ \ ; \ \ \ \ n=0,1,2,...\ .
\end{equation}
In the large-mass $\mu r_{\text{H}}\gg n+1$ regime, the resonance 
spectrum (\ref{Eq23}) can be approximated by the compact analytical
relation
\begin{equation}\label{Eq24}
\alpha_n=\mu^2r^2_{\text{H}}\cdot\Big[1+{{2\sqrt{6}(n+{3\over4})}\over{\mu
r_{\text{H}}}}\Big]\ \ \ \ ; \ \ \ \ n=0,1,2,...\ .
\end{equation}
The discrete resonance spectrum (\ref{Eq24}) of the non-minimal
coupling parameter $\alpha$ characterizes the cloudy
Schwarzschild-black-hole-massive-scalar-field configurations in the
eikonal large-mass regime (\ref{Eq12}).

\section{Summary}

The recently published highly interesting works
\cite{GB1,GB2,ChunHer,Macn} have explicitly proved that, in some
field theories, black holes may support external matter
configurations (hair) made of scalar fields, a phenomenon which is
known by the name black-hole spontaneous scalarization. 
In particular, it has been demonstrated numerically
\cite{GB1,GB2,ChunHer,Macn} that spatially regular (massless as well
as massive) scalar fields with nontrivial couplings to the
Gauss-Bonnet curvature invariant may be supported by central black
holes with regular horizons.

Intriguingly, the numerical results presented in
\cite{GB1,GB2,ChunHer,Macn} have revealed the fact that the
dimensionless physical parameter $\alpha$, which controls the
non-trivial coupling between the Gauss-Bonnet invariant of the
curved spacetime and the supported scalar matter configurations, is
characterized by a discrete resonant spectrum
$\{\alpha_{n}\}_{n=0}^{n=\infty}$ which corresponds to black holes
that support spatially regular nonminimally coupled linearized
scalar field configurations.

In the present paper we have used {\it analytical} techniques in
order to explore the physical properties of the spontaneously
scalarized hairy black-hole spacetimes in the regime of cloudy
(linearized) supported field configurations. In particular, we have derived
the compact WKB analytical formula (\ref{Eq24}) for the discrete 
resonant spectrum which characterizes the non-trivial coupling parameter
$\alpha$ of the composed black-hole-massive-scalar-field theory in
the physical regime $\mu r_{\text{H}}\gg1$ of large field masses.

Finally, it is worth pointing out that one may obtain from the
analytically derived resonance spectrum (\ref{Eq24}) the remarkably
compact formula \cite{Note0n}
\begin{equation}\label{Eq25}
{{\mu r_{\text{H}}}}_{\text{max}}(\alpha)=\sqrt{\alpha+{27\over8}}-\sqrt{{27\over8}}\ \ \ \ \text{for}\ \ \ \ \alpha\gg1\
\end{equation}
for the critical {\it existence-line} which characterizes the hairy
Schwarzschild-black-hole-massive-scalar-field configurations. The
$\alpha$-dependent critical line (\ref{Eq25}) for the masses of the
supported non-minimally coupled scalar fields marks, in the
large-mass $\mu r_{\text{H}}\gg1$ regime, the boundary between bald
Schwarzschild black-hole spacetimes and spontaneously scalarized
hairy black-hole-scalar-field spacetimes. In particular, for a
non-trivial field theory with a given value of the physical coupling
parameter $\alpha$ and for a given mass (radius) of the central
supporting black hole, the hairy
black-hole-nonminimally-coupled-massive-scalar-field configurations
are characterized by the mass inequality ${\mu}(\alpha)\leq
{\mu}_{\text{max}}(\alpha)$.

Interestingly, the {\it analytically} derived formula (\ref{Eq25})
for the critical existence-line of the system implies, in agreement
with the important {\it numerical} results presented in \cite{Macn},
that, for a given value of the dimensionless coupling parameter
$\alpha$, the mass $M$ (horizon radius $r_{\text{H}}$) of the
central supporting black hole is a monotonically decreasing function
of the mass $\mu$ of the nonminimally coupled scalar field.

\bigskip
\noindent
{\bf ACKNOWLEDGMENTS}
\bigskip

This research is supported by the Carmel Science Foundation. I would
like to thank Yael Oren, Arbel M. Ongo, Ayelet B. Lata, and Alona B.
Tea for helpful discussions.



\begin{thebibliography}{99}
\bibitem{Notephi} Here $\phi$ is the non-minimally coupled scalar field.
\bibitem{Noteunits} We shall use natural units in which $8\pi G=c=1$.
\bibitem{Noteho} One may also include a quartic self-interaction term of the form
$-\lambda\phi^4/2$ in the action \cite{Macn}. However, as explicitly
shown in \cite{Macn}, the black-hole-linearized-scalar-field
configurations, which mark the threshold for the black-hole spontaneous 
scalarization phenomenon and are the main focus of the present
paper, are independent of the value of the higher-order parameter $\lambda$.
\bibitem{Notelm} The parameters $\{l,m\}$ (with $l\geq0$ and $-l\leq m\leq l$) in the scalar field decomposition (\ref{Eq6})
are respectively the spherical and azimuthal harmonic indices. For
brevity, we shall henceforth omit these integer indices.
\bibitem{Notemap} Note that the differential relation (\ref{Eq7}) maps
the semi-infinite radial regime $r\in[r_{\text{H}},\infty]$ to the
corresponding infinite regime $y\in[-\infty,\infty]$.
\bibitem{Notephas} The phase shift of ${1\over4}\pi$ in the WKB integral
relation (\ref{Eq13}) reflects the fact that the radially-dependent
potential (\ref{Eq9}) of the composed black-hole-massive-field
system has a classical turning point at $y_-=-\infty$. The
WKB wave field in the classically allowed region $[y_-,y_+]$ should
therefore be matched only once (at the outer classical turning point
$y=y_+$) to the corresponding WKB wave field in the classically
forbidden region $y>y_+$ of the black-hole spacetime. The standard
matching procedure of the wave field across the outer classical
turning point $y=y_+$ [see Eq. (\ref{Eq20}) below] yields the
familiar phase shift of ${1\over4}\pi$ in the second-order WKB
resonance relation (\ref{Eq13}) \cite{WKB1,WKB2,WKB3}.
\bibitem{Notexp} Here we have used the strong inequality
$1/x_+\gg1$ [see Eqs. (\ref{Eq12}) and (\ref{Eq20}), and also Eq.
(\ref{Eq24}) below].
\end{thebibliography}
\end{document}